\documentclass[aps,prd,reprint,groupedaddress]{revtex4-1}
\usepackage{hyperref}
\usepackage{graphicx}
\usepackage{color}
\usepackage{amsmath}
\usepackage{algorithm}
\usepackage{algpseudocode}
\usepackage{amssymb}
\usepackage{lipsum}
\hypersetup{
	pdfnewwindow=true,      
	colorlinks=true,       
	linkcolor=blue,          
	citecolor=blue,        
	filecolor=blue,      
	urlcolor=blue           
}

\begin{document}
	
\title{Optimal Gravitational-wave Follow-up Tiling Strategies Using a Genetic Algorithm}

\author{Nihar Gupte}
\affiliation{Department of Physics, University of Florida, Gainesville, FL 32611, USA}
\author{Imre Bartos}\thanks{E-mail: imrebartos@ufl.edu}
\affiliation{Department of Physics, University of Florida, Gainesville, FL 32611, USA}
	
\begin{abstract}
The identification of electromagnetic emission from gravitational-wave sources typically requires multiple follow-up observations due to the limited fields-of-view of follow-up observatories compared to the poorly localized direction of gravitational waves. Gravitational-wave localization regions are typically covered with multiple telescope pointings using a "honeycomb" structure, which is optimal only on an infinite, flat surface. Here we present a machine-learning algorithm which uses genetic algorithms along with Broyden–Fletcher–Goldfarb–Shanno (BFGS) optimization to find an optimal configuration of tiles to cover the gravitational-wave sky localization area on a spherical surface. 
\end{abstract}
	
\maketitle
	
\section{Introduction}

The LIGO and Virgo observatories now regularly detect gravitational waves from the mergers of black holes and neutron stars \cite{aLIGO2015,Acernese_2014,PhysRevX.9.031040,2017PhRvL.119p1101A}. To learn the most about the discovered events, a vast array of electromagnetic and neutrino observatories carry out follow-up observations to identify multi-messenger emission from the gravitational-wave sources \cite{2017ApJ...848L..12A,2016ApJ...826L..13A,2013CQGra..30l3001B,KohtaBartos}. These include observations in the gamma-ray \cite{2017ApJ...848L..14G,2017ApJ...848L..15S,2016ApJ...826L...6C,2016ApJ...820L..36S}, X-ray \cite{2017Natur.551...71T}, optical \cite{2017Sci...358.1556C,2017ApJ...848L..24V,2017Sci...358.1565E,2017Natur.551...75S,2019ApJ...885L..19C,antier2020grandma,2019ApJ...880L...4H} and radio bands \cite{2017Sci...358.1579H}, as well as high-energy and thermal MeV neutrinos \cite{2016PhRvD..93l2010A,2017PhRvD..96b2005A,2017ApJ...850L..35A,2019ApJ...870..134A}.

Gravitational-wave localization is often limited to tens or even thousands of square degrees \cite{abbott2020gw190425}. While all-sky observatories, such as the Fermi gamma-ray burst monitor \cite{2017PhRvL.119p1101A} and the IceCube Neutrino observatory \cite{2011PhRvL.107y1101B} automatically cover such large sky areas and enable historical searches, the follow-up of poorly localized events is typically more challenging \cite{2019ApJ...880L...4H, ghosh2017hunting, 2019ApJ...885L..19C,2016MNRAS.462.4094S}. Alternative strategies, such as the use of galaxy catalogs, are the most effective for well-localized or nearby events \cite{2015ApJ...801L...1B,2017NatCo...8..831B, antier2020grandma}. 

Optimizing follow-up strategies can substantially improve the chance of discovery \cite{greco2019working, coughlin2019teamwork,ducoin2020optimizing,ghosh2017hunting,2018MNRAS.478..692C,2020MNRAS.492.4768D}. For example, the first optical discovery of the neutron star merger GW170817 was made by a meter-class telescope, Swope, which followed a galactic mass-weighted tiling optimization \cite{2017Sci...358.1556C}. Dedicated follow-up strategies are also particularly important for high-energy gamma-ray follow-up observations, e.g. by the Cherenkov Telescope array, where the source luminosity rapidly decays and the available time window for follow-up is expected to be a few minutes \cite{2018MNRAS.477..639B,2019MNRAS.490.3476B,2014MNRAS.443..738B,2019ICRC...36..790S}, or in some cases a few hours \cite{HESSNature,MAGICNature}.     

In this paper we present a machine learning-based method to optimize the tiling distribution for follow-up observations. Part of our motivation for proposing this method is that it can naturally incorporate the two other key aspects of follow-up: exposure time and scheduling. We base our method on the solution for the {\it optimal equal circle covering problem}. This problem begins with a given polygon and radius, and asks which configuration of circles, all with radius $r$, covers the polygon with the least amount of circles. There are many variants of this problem including covering a polygon with triangles and covering a high-dimensional polytope with with hyper-spheres. Several papers have discussed and even proven optimal circle coverings for triangles, circles, and the sphere \cite{croft2012unsolved, melissen1997packing}. While the dual problem of circle covering, circle packing (finding the optimal way to pack $n$ circles into a polygon), has received considerable attention, solutions to the circle covering problem are less frequent. However, in recent years there has been steady progress in equal circle covering as well as other circle covering problems \cite{nurmela2000covering, melissen1996improved, banhelyi2015optimal, tarnai1995covering, lengyel1996egysegnegyzet, banhelyi2015optimal, stoyan2010covering}. 
	
There are numerous applications of the circle covering problem from telecommunications to land satellite orientations. In telecommunications, it is important to cover polygons, which in this case are cities or countries, with towers so that residents can get full coverage of cell signal. The added advantage of using genetic algorithms and other types of machine learning to solve this problem is the existence of a flexible objective function. Usually, circle covering algorithms deal only with optimally covering a region. It is often the case where we want to assign some benefit for adding extra circles or overlapping at specific regions. Examples of this include low visibility in LIGO skymaps or a probability distribution in the polygon which we are trying to cover. Using genetic algorithms allows domain experts to incorporate the features they are interested in alongside circle covering without much hassle. 
		
Below we present 2 algorithms for gravitational-wave follow-up tiling optimization, which are a special case of equal optimal circle covering. One covers polygons in flat Euclidean space and the other covers spherical polygons on the surface of Earth. While there is no rigorous proof that these are in fact the best coverings, this approach results in strong configurations. To demonstrate this we compare our results with the previous ``honeycomb" method of computing configurations of circle coverings for LIGO sky localizations. Additionally, we provide a ``light" version of the program which runs significantly faster to compute but does not provide the flexibility of genetic algorithms. This is particularly useful for cases where execution time is important. Our code, installation instructions and short documentation can be found at \url{https://github.com/kauii8school/GW-Localization-Tiling}.

Section \ref{sec:method} below presents our tiling method. We present our results in Section \ref{sec:results}. We conclude in \ref{sec:conclusion}.

\section{Method} \label{sec:method}

\subsection{Overview of Genetic Algorithms}

Genetic algorithms (GAs) are a specific type of algorithm from the much broader family of algorithms known as evolutionary computation. GAs and other evolutionary algorithms are usually optimization algorithms inspired by traditional biological evolution. 
    
An ``agent" in GAs represents a possible solution to the problem. In the case of the equal circle covering problem an agent can be fully described by the radius of the circles and a tuple $((x_0, y_0), (x_1, y_1), \dots (x_n, y_n))$ which consists of the centers of the circles. A population is a collection of agents with each individual agent having a different tuple. Initially, the agents are selected randomly which leads to a large search space but as the program begins to converge, the agents are quite similar leading to a small search space likely near the optimum \cite{deb2005population}. The main advantage of a genetic algorithm is having multiple semi-optimal solutions which breed with each other to eventually obtain the strongest traits from each solution. Unlike other optimization techniques, GAs allow jumping from a semi-optimal solution to a completely different yet more optimal solution \cite{flores2016evolutionary}.  This is important for problems like circle covering/packing which have a large (or even continuum) number of optimal solutions \cite{hifi2009beam}.
    
Traditional GAs have 3 steps ``Selection", ``Crossover", and ``Mutation". Selection trims the population to select the strongest individuals for crossover. The ``strength" or ``fitness" of an individual is determined by a user specified objective function, often called a fitness function. Crossover involves combining traits of 2 or more agents to create a new agent. It is often the case that crossover results in a new agent (child) which has a higher fitness than it's predecessors (parents). The final step, ``Mutation", involves slightly perturbing a candidate solution. This is done randomly to members of the population and serves to add additional variance. This way, the program does not get stuck at a local minimum and instead is able to search the full solution space. 
    
For the circle covering problem it is necessary to add an additional step to this traditional method. Since a requirement of the problem is that the solution must cover the polygon, this condition must be enforced. One way to do so is to incorporate a penalty for the amount of uncovered area into the fitness function. This way, agents which fail to cover the region will be gradually removed leaving only agents which cover the region. The problem with this approach is that it throws away too many good agents and often times will not converge. For example, a near optimum solution of covering may miss only a small bit of the region and will be thrown out simply because it does not cover the entire region. Many times it is the case that a slight movement of a circle in the agent results in a full cover. Therefore, we follow the clever approach of \cite{flores2016evolutionary} and introduce a ``repair function". In essence this function fixes agents to cover the region as best as possible. The program finishes when there are no longer enough agents in the population to breed or the number of desired generations is reached. The only difference between the flat algorithm for Euclidean space and the spherical algorithm is the methods for computing geometrical operations (intersection, area, union, alphashape, Voronoi diagrams, etc.). In particular for computing the configurations for localization maps, first the map needs to be separated out into non-connected regions (since localization maps aren't always one polygon). Then the border of the maps must be found. In this code we employ DBSCAN for clustering and alphashape for finding the border of the maps. Thus the pseudo-code for the full algorithm is as follows:
    
\bigskip
    
\begin{algorithm}[H]
 \caption{Genetic Algorithm for Circle Covering}\label{GA}
 \begin{algorithmic}
    \Procedure{GA}{$a,b$}\Comment{Optimal circle layout given region}
    \State \texttt{agent\_list = initialize\_agents(region)}
     \For{\texttt{i in num\_generations}}
        \State \texttt{repair\_agents(agent\_list)}
        \State \texttt{determine\_fitness(agent\_list)}
        \State \texttt{select(agent\_list)}
        \State \texttt{crossover(agent\_list)}
        \State \texttt{mutate(agent\_list)}
     \EndFor
     \State \textbf{return} \texttt{best\_agent(agent\_list)}
    \EndProcedure
 \end{algorithmic}
\end{algorithm}
    
\subsection{Initialization and Repair function}
    
An agent is initialized by randomly generating an initial guess of circle centers within the polygon region. To create the first generation we need to supply an initial guess of how many circles the final solution will have. 
%
We adopt the strategy employed in \cite{nurmela2000covering}, which computationally solved minimal radius circle covering. That is we use the Broyden–Fletcher–Goldfarb–Shanno (BFGS) algorithm to find layouts of circles which maximize the area of the region covered. BFGS is a quasi-Newtonian method which iteratively finds a stationary point. The main advantage of BFGS over other optimization algorithms is that the exact Hessian matrix is not needed and instead can be approximated. This is valuable since finding the Hessian for non-linear problems like circle covering is often non-trivial \cite{avriel2003nonlinear}. The required input for BFGS optimization is an initial guess tuple $(x_1, x_2, \dots x_n)$ and a function. In this case, the tuple is a flattened version of the centers of the agent $(x_0, y_0, x_1, y_1, \dots x_{n}, y_n)$ and the function returns the area of intersection between the agent and the region. Additionally, methods for calculating the first and second derivatives of the function can be supplied to BFGS for speedup though this is not required.
    
\begin{figure}[h]
    \hspace*{-.75cm}
    \includegraphics[width=10cm]{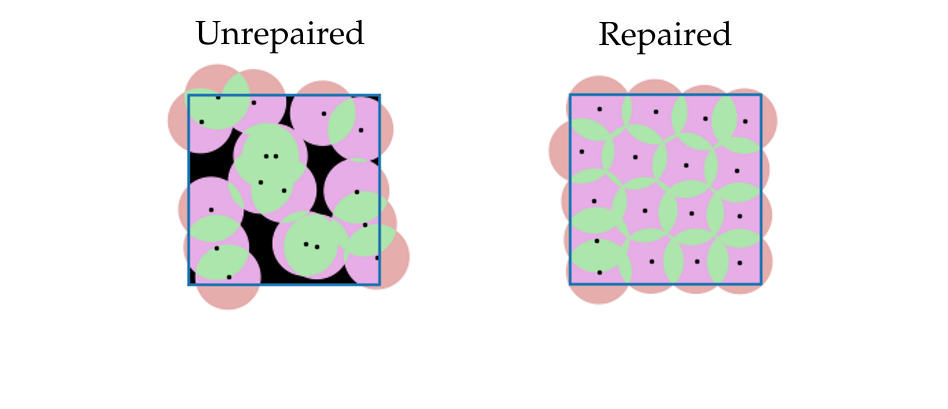}
    \caption{Image of un-repaired agent (left) attempting to cover a square region which is then repaired by BFGS (right). Red is the intersection between the circles and outside the region, purple is the intersection between circles and the inside the region, green is the self-intersection between the circles, and black is the uncovered area. Additionally, the black dots represent the centers of circles. This agent has 17 circles which is not the optimal number of circles for covering this square region. This is intended as BFGS does not find the optimal layout for covering, but rather finds a generic covering. Also note most agents will not require this much repairing, it is only for the first generation that a large repair is needed. }
    \label{Flat}
\end{figure}

After BFGS is applied, we test that the agent is indeed covering the region. However, not every agent is repairable. In fact many will not be, for example, if the optimal solution for a cover is 15 circles a layout with 14 circles will never cover the region. Additionally, sometimes, even a layout with 15 circles may not cover the region after BFGS is applied. The agents which are not repairable are discarded. If a sufficient starting population is given this is not an issue. Finally, we remove circles which fully overlap with the agent and are redundant. See figure \ref{Flat} to see a before and after of BFGS optimization on flat Euclidean space and Earth for an initial agent.

\subsection{Fitness and Selection}
    
The fitness function in this paper is determined by 3 factors: area of intersection between the circles and the outside of the region, area of self-intersection between the circles, and the number of circles. These factors are normalized to be between $0$ and $1$ so that it is easier to fine tune the fitness function. For a given agent, let $\Omega \equiv \frac{\text{area of intersection outside the region}}{\text{area of region}}$, let $\Gamma \equiv \frac{\text{area of self intersection of circles}}{\text{total area of all circles}}$, and $N \equiv \frac{\text{number of circles}}{\text{initial number of circles}}$. Then the fitness function is: 
$$
\text{fitness(agent)} =  (\alpha + \beta + \gamma) - \alpha \Gamma - \beta \Omega - \gamma N
$$
Here $\alpha, \beta$ and $\gamma$ are constants which determine the importance of each factor. As stated before, there is a degree of freedom when it come to the configurations. These parameters allow the user to reduce this degree of freedom. For example, if the least amount of self-intersection is desired, $\Gamma$ can be set to a very low number or even 0. The worst possible fitness is $0$ and the best is $\alpha + \beta + \gamma$. In the flat algorithm $\alpha = 2.1$, $\beta = .8$, $\gamma = .5$. In the spherical algorithm $\alpha = 2.5$, $\beta = 1.1$, $\gamma = .7$. Note that this fitness function can be easily edited by others to incorporate the features they are interested in. 
    
As for selection the bottom 20\% of agents are deleted from the population. This number can be changed depending on the size of the initial population and the desired rate of convergence. I.e. higher initial populations can have higher deletion rates. 
    
\subsection{Crossover}

In GAs there are multiple ways to construct crossover functions which vary on a problem by problem basis. In this case it is necessary to need to breed agents so that their desirable traits are conserved. Thus it does not make sense to randomly select circles from both of the parents and add them to the child as it will result in a configuration which is not similar to either parent. In particular, if a crossover function is not selected carefully, a consequence could be having all the circles on only half the region. Furthermore, this configuration will likely be irreparable. In essence, the aim is to construct a child which takes circles of parents which are locally close and selects one of them. In practice, this is done via Voronoi diagrams. Given $(p_1, p_2, \dots p_n)$ points (sometimes called seeds), a Voronoi diagram on the plane is a partition of the plane into $n$ sub-regions such that each sub-region $S_i$ consists of points closer to $p_i$ than any other point. 
    
\begin{figure*}[t!]
    \includegraphics[width=\textwidth]{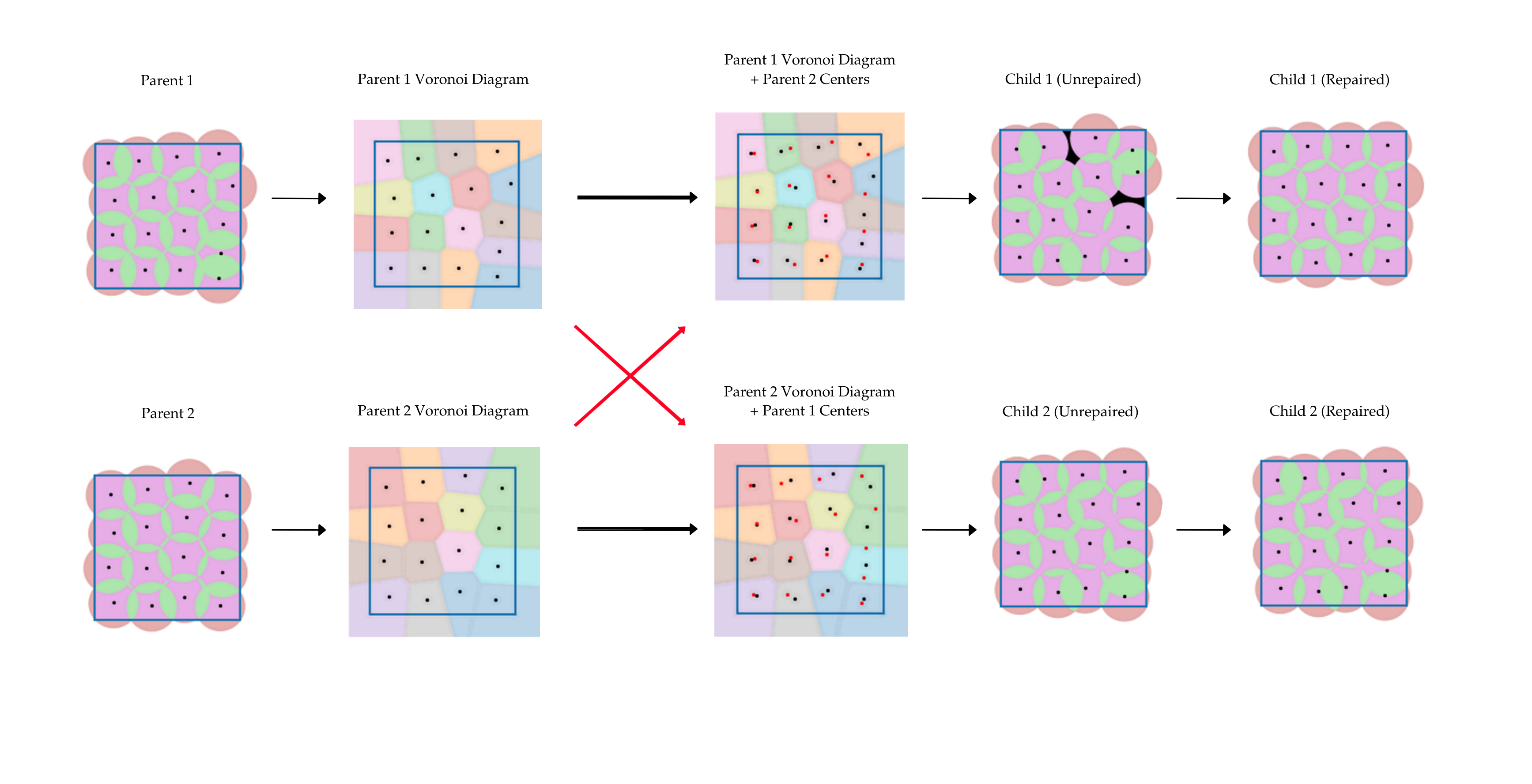}
    \caption{Crossover function for the genetic algorithm which utilizes Voronoi diagrams. The top figure describes the creation of child 1 and the bottom describes the creation of child 2. From left to right, first two parents are selected. Then the Voronoi diagrams for each parent are computed. Then the points of the opposing parent (they are in red here) are placed onto the Voronoi diagram. We then iterate through each of the partitions randomly and exclusively select either all the red points in the partition or the black point. Then these points are appended to child 1 and circles are generated. Finally, the child is repaired using BFGS in case crossover caused gaps to form.}
    \label{Voronoi}
\end{figure*}

The way the breeding works is by first constructing Voronoi diagrams for both parents. Then consider the Voronoi diagram from parent 1. Stack onto it the center points of parent 2. Iterate through the regions of the Voronoi diagram. If the region does not contain a point from parent 2, append to the circle list of child 1, the point which generated the current region. If instead the region contains points from parent 2 randomly select either all the points from parent 2 contained in the region, or the point from parent 1. Append these points to the circle list of child 1.     The only difference in the spherical algorithm is the use of a spherical Voronoi function \cite{de2009robust}.  See figure \ref{Voronoi} for a pictorial description.

In this case we iterate through half the length of the population, randomly select 2 agents and breed them. Each bred pair produces 2 children. After iterating through the entire population the children and parents are added back into population and repaired as indicated in Algorithm \ref{GA}. 
    
\subsection{Mutation}
    
The final step in the algorithm is to ``mutate" random agents in the population. There are 2 types of mutation, removing circles and moving circles. It isn't effective to just remove random circles from the agent. Rather, it is better to remove the worst circles. There are 2 types of ``worst" circles, circles which intersect the region the least and circles which self intersect the most. By self intersect the most, we mean the circle with the highest intersection between the circle list of the agent. The second type of mutation is moving a random circle by a random specified amount. The mutation rate for removal is .1 where as the mutation rate for moving the circle is $0.2$. Again, these mutation rates can be changed based on the size of the initial population.

\section{Results} \label{sec:results}

\begin{figure}[h]
    \includegraphics[width=0.5\textwidth]{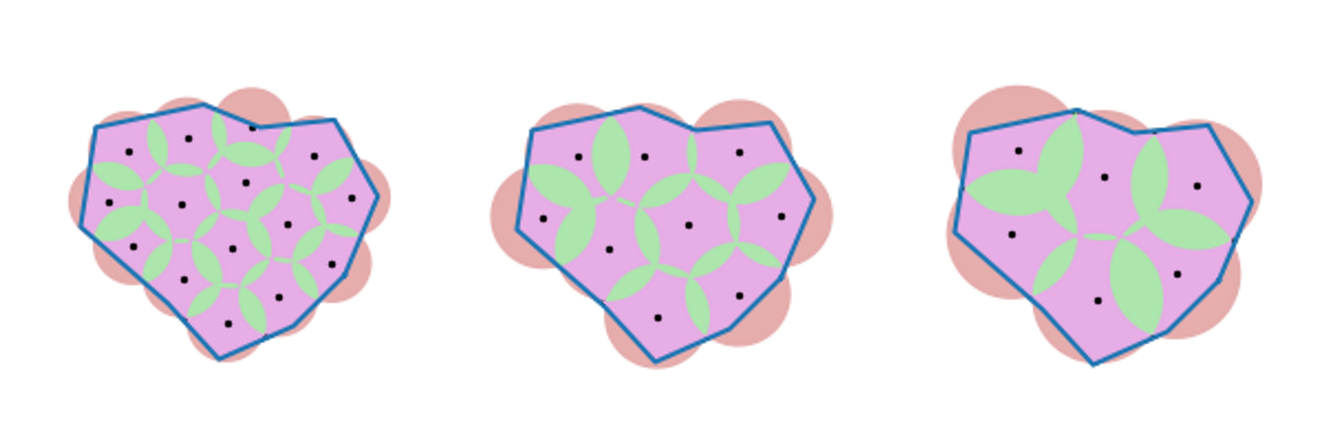}
    \caption{Three different coverings of a randomly generated irregular polygon by genetic algorithms. From left to right, the radii of the circles are 1.5, 2, and 2.5. Red is the intersection between the circles and outside the region, purple is the intersection between circles and the inside the region, green is the self-intersection between the circles, and black is the uncovered area. Additionally, the black dots represent the centers of circles.}
    \label{flat_results}
\end{figure}

\begin{figure*}[h]
      \begin{minipage}[h]{1.0\linewidth}
         \centering
         \includegraphics[width=0.75\linewidth]{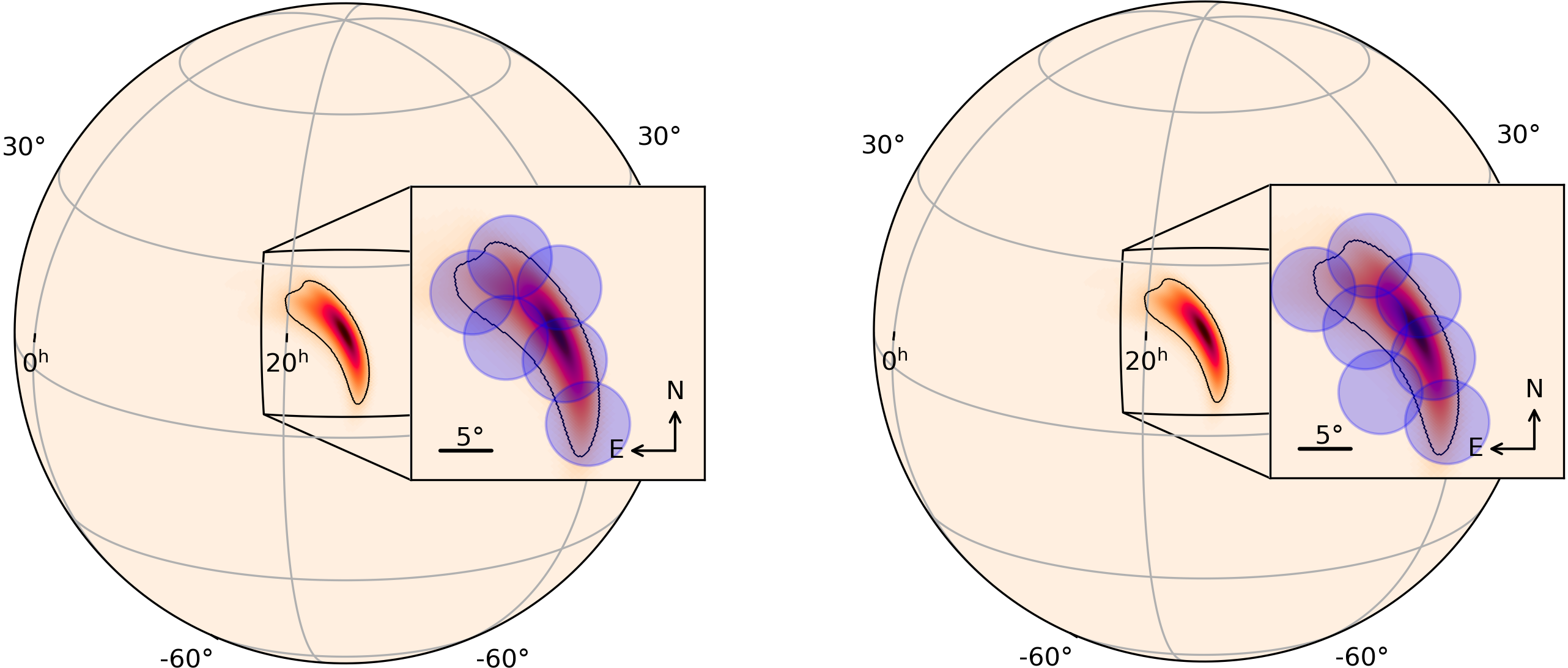}
         \vspace{5.00mm}
      \end{minipage}
     \begin{minipage}[h]{1.0\linewidth}
        \centering
        \includegraphics[width=0.75\linewidth]{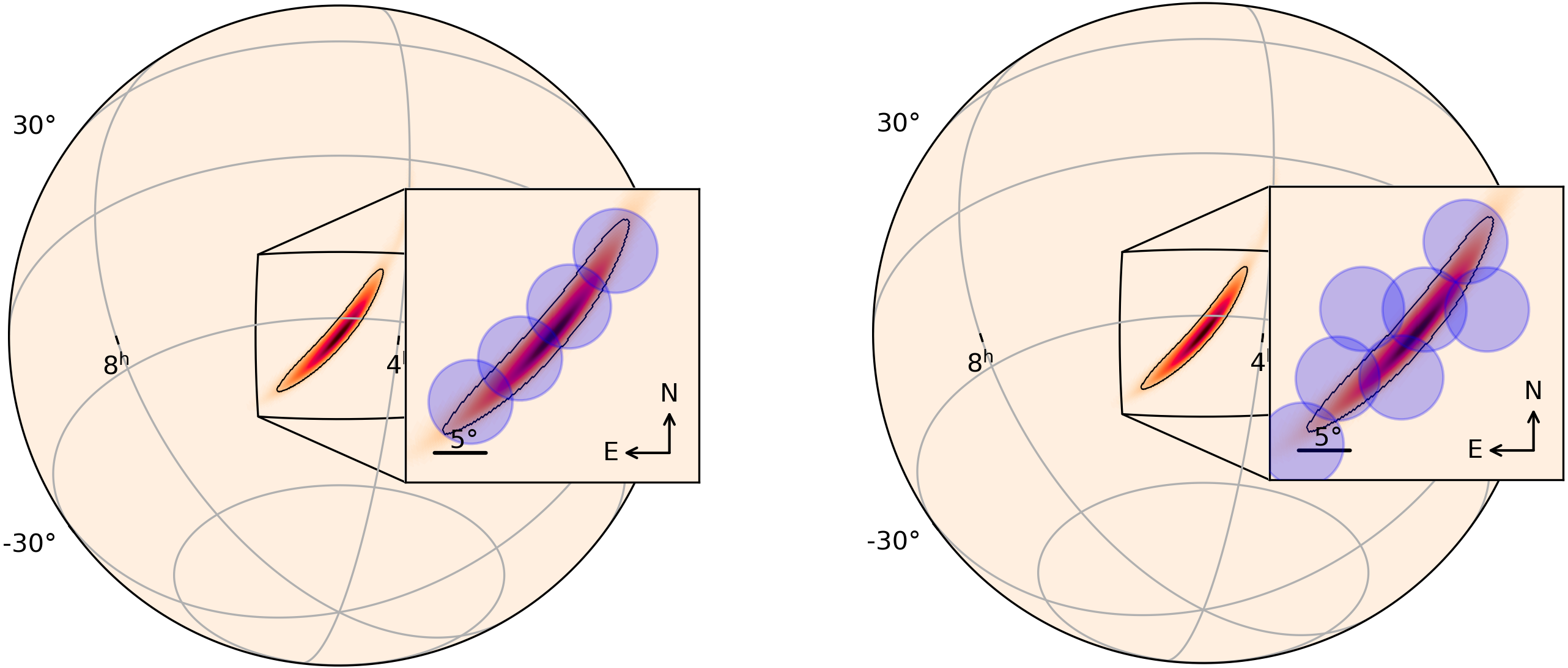}
      \end{minipage}    
    \caption{Comparisons between genetic algorithm (left) and honeycomb method (right) for covering LIGO sky localizations. The black line around the sky localizations show their 90\% credible region. For both examples genetic optimization results in less required tiling.}
    \label{ligo_results}
\end{figure*}

The algorithm presented above performs strongly given most polygons on flat space. Presented in figure \ref{flat_results} are a few configurations for 3 different radii for a randomly generated irregular polygon. Additionally, Fig.  \ref{ligo_results} compare the GA to the honeycomb method from \cite{2019MNRAS.490.3476B}. We also computed the tiling for 100 simulated LIGO sky localizations (obtained from \cite{2019MNRAS.488.4459C}) with both the hexagonal method and genetic algorithm method. The genetic algorithm performed better in 80\% of the cases, while in the other 20\% the genetic algorithm performed equally well. Overall, for skymaps requiring at least 10 tilings, the genetic algorithm presented a $30\%$ reduction in the number of tiles compared to the honeycomb method.

\section{Conclusion} \label{sec:conclusion}

We developed an optimization strategy for the distribution of tilings in gravitational-wave follow-up observations using a genetic algorithm and a BFGS repair function. Compared to the standard ``honeycomb" tiling strategy that is optimal in infinite flat surfaces, our method is superior in 80\% of the cases, and in some cases it performs substantially better (e.g., Fig. \ref{ligo_results}).

We made the optimization algorithm available as a python code at \url{https://github.com/kauii8school/GW-Localization-Tiling}. This Github page contains three codes, Sphere\_Light.py, Flat\_GA.py, and Sphere\_GA.py which correspond to the light version of the code, the Euclidean genetic algorithm, and the Spherical genetic algorithm respectively. The light version of the code is not a genetic algorithm but rather repeatedly uses BFGS optimization to try and cover a region. Specifically, if the user specifies a range of number of circles the program will start at the lower bound and try to cover the region. If it is unable to do so it will add more circles until either the upper bound is reached or the region is covered. If the program is able to cover the region this configuration will be considered as an ``optimal covering". 

The optimization strategy developed here is applicable beyond gravitational-wave follow-up. It is a generalized solution for the equal circle covering problem on a spherical surface, and to higher dimensional polytopes. We also note that this algorithm can be easily extended to optimize for a range of circle radii rather than just a uniform radius. 

For the optimization of tiling, we focused on a machine learning-based algorithm as they can also naturally incorporate other aspects of follow-up optimization, such as galaxy distributions, the temporal variations of the sources luminosity, and the follow-up observatories direction-dependent sensitivity. These elements are key in determining the best follow-up strategy. Our next step will be to generalize the algorithm presented here to account for these factors. 

\begin{acknowledgments}
The authors thank Tristano Di Girolamo and Ik Siong Heng for their useful feedback, and the University of Florida for their generous support.
\end{acknowledgments}

\bibliographystyle{apsrev}
\bibliography{Refs}


\end{document}